\documentclass[11pt]{article}
\usepackage{latexsym,cite}
\usepackage{graphicx}
\usepackage{dcolumn}
\usepackage{bm}
\usepackage{xcolor}
\usepackage{float}
\usepackage{epsfig}
\input amssym.def
\input amssym.tex

\newcommand{\be}{\begin{equation}}
\newcommand{\ee}{\end{equation} }
\newcommand{\beqa}{\begin{eqnarray}}
\newcommand{\eeqa}{\end{eqnarray} }
\newcommand{\ba}{\begin{array}}
\newcommand{\ea}{\end{array}}
\newcommand\dis{\displaystyle}

\newcommand{\half}{{{\textstyle\frac{1}{2}}}}

\newcommand\cP{{\cal P}}

\newcommand\cT{{\cal T}}
\newcommand\cV{{\cal V}}

\newcommand\vn{\vec{n}}

\newcommand\kB{k_{{\scriptscriptstyle{\rm B}}}}

\newcommand\conti{{{\scriptscriptstyle{\rm conti.\,approx.}}}}
\newcommand\superc{{{\scriptscriptstyle{\rm supercool}}}}
\newcommand\superh{{{\scriptscriptstyle{\rm superheat}}}}

\begin{document}
\begin{titlepage}
\title{\vskip -110pt
\vskip 2cm Existence of a critical point in the phase diagram  of the  ideal relativistic neutral  Bose gas\\}
\author{\sc      Jeong-Hyuck Park${}^{\dagger\ast}$  ~and~ Sang-Woo Kim${}^{\natural}$}
\date{}
\maketitle \vspace{-1.0cm}
\begin{center}
~~~\\
${}^{\dagger}$Department of Physics, Sogang University\\ Shinsu-dong, Mapo-gu, Seoul 121-742, Korea\\
~~~\\
${}^{\natural}$High Energy Accelerator Research Organization (KEK)\\ Tsukuba, Ibaraki 305-0801, Japan\\ 
~~~\\
~~~\\
\end{center}
\begin{abstract}
\vskip0.1cm
\noindent
We explore the phase transitions of  the  ideal relativistic neutral Bose gas  confined in a cubic box, without assuming  the thermodynamic limit  nor  continuous approximation.  While the corresponding  non-relativistic canonical partition function is essentially a one-variable  function depending on a particular  combination of temperature and volume, the relativistic canonical partition function is genuinely a two-variable  function of them.  Based on an   exact expression of  the  canonical partition function, we performed   numerical computations for up to $10^{5}$ particles.   We report that   if the number of particles is equal to or greater than a critical value, which amounts  to  ${7616}$,  the ideal  relativistic neutral   Bose gas  features   a     spinodal curve  with a critical point. This enables us to depict  the phase diagram of the ideal Bose gas. The consequent  phase transition  is  first-order below the critical pressure  or second-order at the critical    pressure. The   exponents  corresponding to the singularities are   $1/2$  and $2/3$  respectively. We also verify the recently observed `Widom line' in the supercritical region. 
\end{abstract}
{\small
\begin{flushleft}
~~~~~~~~\textit{PACS}:  03.75.Hh,  51.30.+i
~~~~~~~${}^{\ast}$\textit{Corresponding electronic address}: park@sogang.ac.kr\\
\end{flushleft}}
\thispagestyle{empty}
\end{titlepage}
\newpage

\section{Introduction}
Spinodal curve, by definition,  consists of points in a phase diagram where the isothermal volume derivative of pressure  vanishes~\cite{spinodalnuclear,Sasaki:2007qh}:
\be
\partial_{V}P(T,V)=0\,.
\label{defspinodal}
\ee
The curve corresponds to  the border line between thermodynamic stable  and unstable  regions, ${\partial_{V}P(T,V)<0}$ and ${\partial_{V}P(T,V)>0}$ respectively~\cite{Huang}.  Moreover, if it exists,  the spinodal  curve  amounts to a first-order phase transition under constant pressure.   The temperature derivative at fixed pressure
acting on an arbitrary physical quantity which is a function    of temperature  and volume, is given by   the chain rule of calculus:
\be
\textstyle{
\left.\frac{\partial~}{\partial T}\right|_{P}=
\left.\frac{\partial~}{\partial T}\right|_{V}
-\left[\frac{\partial_{T}P(T,V)}{\partial_{V}P(T,V)}\right]\!
\left.\frac{\partial~}{\partial V}\right|_{T}\,.}
\label{pdTP}
\ee
On the spinodal curve, the denominator of the second term vanishes, and   the temperature derivative  of a generic physical quantity along the isobar  diverges~\cite{7616Park}.    This provides a possible mechanism  for  a finite system  to manifest  genuine mathematical singularities, without taking the thermodynamic limit~\cite{LeeYang}. \\

\noindent If we fill a rigid box with water to  full capacity and heat the box,  the temperature will increase but  hardly  the water evaporates. According to a well-known  standard argument  against the emergence of  a singularity from a finite system that  is based on the analytic property of the canonical partition function~\cite{Huang}, no  discontinuous  phase transition should occur.   Nevertheless,   opening   the lid  will set the pressure as constant  or at {$1$ atm}, and the water will surely   start to boil at $100$ degree Celsius.  The point is that   the usual  finiteness of physical quantities in a canonical ensemble is  for the case   of keeping  the volume fixed. Once we switch to the alternative constraint of  keeping the pressure constant,    first order phase transition featuring genuine mathematical  singularities  may arise from   a system possessing  finite number of  physical degrees. The question is then the existence of the spinodal curve. \\

\noindent In our previous work~\cite{7616Park},  we   investigated    the thermodynamic instability of the  ideal non-relativistic    Bose gas confined in a cubic box,    based on an   exact expression of  the corresponding canonical partition function. The result was that   if the number of particles is equal to or greater than a certain critical value\footnote{The number $7616$  can be regarded as  a characteristic number of a cube.  For a sphere we get the critical number,  $10458$.} which turns  out to be  ${7616}$,  the ideal Bose gas  subject to  Dirichlet boundary condition  indeed  reveals  thermodynamic instability characterized  by a pair of spinodal curves   and consequently  undergoes   a first-order phase transition under constant pressure. The two spinodal curves  are identified as   the supercooling and the  superheating of the ideal Bose gas,
\be
\ba{l}
\kB T_{\superc}= \tau_{P}^{\ast}\times\textstyle{[\pi^{2}\hbar^{2}}/(2m)]^{3/5}P^{2/5}\,,\\
\kB T_{\superh}= \tau_{P}^{\ast\ast}\times\textstyle{[\pi^{2}\hbar^{2}}/(2m)]^{3/5}P^{2/5}\,,
\ea
\label{supert2}
\ee
where  $\kB$ is the Boltzmann constant,  $m$ is the mass of the particle, $P$ is an arbitrary given pressure, and   $\tau_{P}^{\ast}$, $\tau_{P}^{\ast\ast}$ are dimensionless constants of which the numerical values depend  on the the number of particles, $N$,  as follows~\cite{7616Park}:
\begin{center}
\begin{tabular}{c|l|l}
{$N$}&$~~~~~~\tau_{P}^{\ast}$  &~~~~~$\tau_{P}^{\ast\ast}$  \\
\hline
$7616~~$&$~~1.054\,369\,4113~~$&$~~1.054\,369\,4116~~$\\
$10^{4}\,$&$~~1.052\,70~~$&$~~1.052\,77~~$\\
$10^{5}\,$&$~~1.0410~~$&$~~1.0424~~$\\
$10^{6}\,$&$~~1.034~~$&$~~1.036~~$
\end{tabular}
\end{center}
~\newline

Between the two temperatures on the isobar, every physical quantity  zigzags  or becomes triple valued, implying  the existence  of  three  different phases   during the discontinuous  phase transition.  This includes the volume as well as the number of particles occupying the ground state. Hence, the corresponding first-order phase transition coincides with the Bose-Einstein condensation (BEC) in both  the  coordinate and the momentum spaces.   This result is comparable to the well known     BEC  temperature  of  the  ideal non-relativistic  Bose gas in the thermodynamic limit: 
\be
\textstyle{\kB T_{\conti}=\textstyle{
\textstyle{[2\pi\hbar^{2}/m]^{3/5}[P/\zeta(5/2)]^{2/5}}
\simeq 1.0278\!
\times\!\textstyle{[\pi^{2}\hbar^{2}}/(2m)]^{3/5}P^{2/5}\,,}}
\label{BECt}
\ee
where $\zeta$ is  Riemann zeta function.  In  standard textbooks, 
\textit{e.g.~}\cite{Huang,LandauPart1,Kadanoff,Pethick,Sachs},  
this  temperature is typically obtained   by treating the ground state separately and applying the continuous  approximation to all the rest of the excited states.  
From the comparison of Eq.(\ref{supert2}) and Eq.(\ref{BECt}),  it appears that   the continuum limit  smoothes out some details of the thermodynamic quantum   structure.\\


\noindent In the present paper, we generalize our previous work to the relativistic case. We investigate  into the  spinodal curve  of ideal relativistic  quantum gas  of neutral bosons  confined in a cubic box, subject to Dirichlet boundary condition.  While the  non-relativistic canonical partition function is essentially a one-variable  function depending on a particular  combination of temperature and volume, the relativistic canonical partition function is genuinely a two-variable  function of them.   We show that   \textit{if the number of particles is equal to or greater than the critical number ${7616}$, the ideal  relativistic  neutral Bose gas  features   a     spinodal curve  with a critical point where the supercooling and the superheating lines converge.}  Further,  we  verify the recently observed  `Widom line'~\cite{Widom} emanating from  the critical point   as the crossover between liquid-like and gas-like behaviour in  the supercritical region~\cite{NPWidom1} (see also \cite{NPWidom2} and \cite{WidomXu}).  The critical number coincides with that of the non-relativistic system and hence, the present paper  confirms  our previous numerical results~\cite{7616Park}.  \\

\noindent The main motivation for us to consider the relativistic ideal Bose gas in the present paper as a  generalization of the   non-relativistic system we studied previously is,  by letting  the  corresponding  canonical partition function be  a two-variable  function, to observe  a more intriguing phase diagram such as a critical point and the Widom line.  After all, quantum mechanics and relativity are two cornerstones  of modern physics.     For pioneering earlier works   on the ideal relativistic Bose gas  —\textit{via}   the continuous  approximation — we refer  to 
Refs.\cite{relativisticBEC1,relativisticBEC2,relativisticBEC3,relativisticBEC4}. For recent discussion see \cite{relativisticBEC5} and references therein.  In the present  paper, we restrict  ourselves to neutral bosons. The inclusion of anti-bosons, as in\cite{antiboson1,antiboson2,antiboson3}, will be treated separately elsewhere~\cite{preparation}.   \\

\noindent The rest of the paper is organized as follows:\\
In section~\ref{secGeneral},  we present model independent analysis   on canonical ensemble. We first review  three different expressions  for the canonical partition function of  generic non-interacting identical bosonic particles. We then, simply by assuming the existence of a spinodal curve,  show that  under constant pressure 
first-order phase transitions feature  singularities  of exponent $1/2$ while   second-order phase transitions have the critical exponent $2/3$.\\
Section~\ref{secBose} is exclusively devoted to  the ideal relativistic neutral Bose gas confined in a cubic box.  After analysing non-relativistic and ultra-relativistic   limits, we present our numerical results. When the number of particles is equal to or greater than ${7616}$, the ideal  relativistic  Bose gas  reveals   a   spinodal curve  with a critical point. We draw the corresponding phase diagram. \\
The final section~\ref{secDISCUSSION} summarizes  our results with  comments.\\

\section{General analysis\label{secGeneral}}

\subsection{The canonical partition function: review}
When a single particle system is completely solvable,  each quantum state  is uniquely specified by a set of good quantum numbers,  which we simply denote here   by a vector notation, $\vec{n}$. With the corresponding energy eigenvalue $E_{\vn}$ and ${\beta=1/(\kB T) }$,  we define  for   each  positive integer $a$,
\be
\textstyle{\lambda_{a}:=\sum_{\vec{n}\,}e^{-a\beta E_{\vec{n}}}}\,,
\label{lambda}
\ee
where the sum is over all the quantum states. In particular, 
when ${a=1}$, $\lambda_{a}$ coincides with  the canonical partition function of the single particle.\\

\noindent For an  $N$-body system composed of  the  non-interacting identical bosonic particles  above, we may write the corresponding canonical particle function in three different manners:
\begin{itemize}
\item As first obtained  by  Matsubara~\cite{Matsubara}  and Feynman~\cite{Feynman},
\be
\textstyle{{Z}_{N}=\sum_{{{m_{a}}}}
\prod_{a=1}^{N}{(\lambda_{a})^{m_{a}}}/{(m_{a}!\,a^{m_{a}})}\,,}
\label{ZNFeynman}
\ee
where the sum  is over all the partitions of   $N$, given by non-negative integers  $m_{a}$ with  \small{$a=1,2,\cdots, N$} satisfying   $N=\sum_{a=1}^{N} a\,m_{a}\,$.  

\item By a recurrence relation first  derived by Landsberg~\cite{Landsberg},
\be
\textstyle{Z_{N}=\left(\sum_{k=1}^{N}\lambda_{k}Z_{N-k}\right)/N\,,}
\label{recZ}
\ee
where we set ${Z_{0}=1}$. 

\item From Ref.\cite{7616Park},
\be
\textstyle{Z_{N}=\det(\Omega_{N})\,(Z_{1})^{N}/N!}
\label{ZNPsi}
\ee
where $\Omega_{N}$  is an almost triangularized   $N{\times N}$ matrix  the entries  of which  are  defined  by
\[
\Omega_{N}[a,b]:=\left\{\ba{cl}
\lambda_{a{-b}{+1}}{/\lambda_{1}}~~&~~\mbox{for~~~} {b\leq a}\\
-a{/\lambda_{1}}~~&~~\mbox{for~~~} {b=a+1}\\
0~~&~~\mbox{otherwise}\,.\ea\right.
\]
In particular, every diagonal entry is unity. 
\end{itemize}

The first expression (\ref{ZNFeynman}) implies that, compared to ideal Boltzmann gas, ideal Bose gas has higher probability for the particles to occupy the same quantum state~\cite{7616Park}.  The last expression (\ref{ZNPsi}) is useful for us to see when the canonical partition function reduces to the  conventional approximation~\cite{Gibbs}:
 \be
 {Z_{N}\longrightarrow (Z_{1})^{N}/N!}
 \label{conapp}
 \ee
which  would be only valid if all the particles  occupied  distinct states, as in  high temperature limit. Indeed, in this limit we have  $\lambda_{1}\rightarrow\infty$. Hence  $\det(\Omega_{N})\rightarrow 1$  and the reduction (\ref{conapp}) holds. \\

According to the Hardy-Ramanujan's estimation, the number of  possible partitions  grows  exponentially like $\textstyle{e^{\pi\sqrt{2N/3}}/(4\sqrt{3}N)}$, and this would make any numerical computation based on the expression  (\ref{ZNFeynman}) practically hard for large $N$.  Among the three above, the recurrence relation (\ref{recZ}) provides the most efficient  scheme of  $N^{2}$ order computation.

\subsection{Critical and noncritical exponents: universal results\label{Secexponent}} 
When the energy eigenvalue of the single particle system $E_{\vn}$ (\ref{lambda}) depends on volume, the canonical partition function depends on temperature and volume, so does the pressure:
\be
P(T,V)=\kB T\partial_{V}\ln Z_{N}(T,V)\,.
\ee
We   consider, at least locally,\footnote{Since the vanishing of $\partial_{V}T(P,V)$  implies the divergence of  $\partial_{T}V(P,T)$, inverting  $P(T,V)$ to express the volume  as a function of $P$ and $T$ is not conceivable  close to the spinodal curve, as we see shortly  from the relation (\ref{useful}).}  inverting $P(T,V)$ to express the temperature as a function of $P$ and $V$ like $T(P,V)$.
Plugging this expression into $\partial_{V}P(T,V)$, let us define a quantity $\Phi$ as a function of $P$ and $V$:
\be
\Phi(P,V):=-\left.\partial_{V}P(T^{\prime},V)\right|_{T^{\prime}\rightarrow T(P,V)}\,.
\label{Phi}
\ee
As we change  the volume  keeping the pressure  constant, $\Phi$ indicates whether the thermodynamic instability develops or not on the isobar.     With the minus sign in front, now  negative $\Phi$ corresponds to the thermodynamic instability.

In a similar fashion to  (\ref{pdTP}) we write
\be
\textstyle{
\left.\frac{\partial~}{\partial V}\right|_{P}=\left.\frac{\partial~}{\partial V}\right|_{T}
+\Phi\,\partial_{P}T(P,V)\left.\frac{\partial~}{\partial T}\right|_{V}\,,}
\ee
from which it is straightforward to obtain
\be
\textstyle{\partial_{V}\Phi(P,V)=-\partial^{2}_{V}P(T,V)-\Phi\,\partial_{P}T(P,V)\,\partial_{T}\partial_{V}P(T,V)\,,}
\label{dVP}
\ee
and relate the first and second volume derivatives of $T(P,V)$ to  $\Phi(P,V)$ as
\be
\ba{l}
\partial_{V}T(P,V)=\Phi\,\partial_{P}T(P,V)\,,\\
\partial^{2}_{V}T(P,V)=\partial_{V}\Phi(P,V)\,\partial_{P}T(P,V)+
\Phi\,\partial_{P}\partial_{V}T(P,V)\,.
\ea
\label{useful}
\ee
Further, on the spinodal curve (${\Phi=0}$) -- if it exists --  we have
\be
\textstyle{
\left.\frac{{\rm d} T}{{\rm d}V}\right|_{\Phi=0}=-\frac{\partial^{2}_{V}P(T,V)}{\partial_{T}\partial_{V}P(T,V)}\,.}
\label{ForMax}
\ee

Now critical point can be  defined as follows.  At the critical pressure ${P=P_{c}}\,$, the system is  generically stable  \textit{i.e.~}$\Phi>0$, except at only one point  that is the critical point.  This implies that the critical pressure line is tangent to the spinodal curve, such that  at the critical point,    $\Phi$ and $ \partial_{V}\Phi$  vanish:
\be
\ba{cc}
\Phi(P_{c},V_{c})=0\,,~~~~&~~~~\partial_{V}\Phi(P_{c},V_{c})=0\,.
\ea
\label{criticaldef}
\ee
Moreover,  from  (\ref{Phi}), (\ref{dVP}), (\ref{useful}), these two conditions are equivalent to
\be
\ba{cc}
\partial_{V}P(T_{c},V_{c})=0\,,~~~~&~~~~\partial^{2}_{V}P(T_{c},V_{c})=0\,,
\ea
\label{criticalprop1}
\ee
and also to
\be
\ba{cc}
\partial_{V}T(P_{c},V_{c})=0\,,~~~~&~~~~\partial^{2}_{V}T(P_{c},V_{c})=0\,.
\ea
\label{criticalprop2}
\ee
Eq.(\ref{criticalprop1}) corresponds to the usual definition of the critical point as an
inflection point in  the critical isotherm on a $(P,V)$ plane~\cite{Huang}.  With (\ref{ForMax}), Eq.(\ref{criticalprop1})  implies that the critical point is an extremal point on the spinodal curve.  On the other hand, Eq.(\ref{criticalprop2}) implies  that the expansion of $T(P,V)$ near the critical point  on the critical isobar  starts from the cubic order in $V-V_{c}$ (for related earlier works see \cite{Sengers} and references therein):
\be
T(P_{c},V)-T_{c}=\textstyle{\frac{1}{6}}(V-V_{c})^{3}\partial^{3}_{V}T(P_{c},V_{c})\,+\,\mbox{higher~orders}\,.
\ee
Clearly this leads to the following  critical exponent:
\be
\ba{ll}
V/V_{c}-1~\sim~{\left|{T/T_{c}-1}\right|}^{\beta}\,,~~~&~~~\beta=1/3\,.
\ea
\label{beta}
\ee
Similarly, from (\ref{dVP}) and (\ref{criticalprop1}),    the expansion of $\Phi$ on the critical isobar starts from the quadratic order in $V-V_{c}$ such that
\be
\Phi(P_{c},V)~~\sim~~(V/V_{c}-1)^{2}~~\sim~~{\left|{T/T_{c}-1}\right|}^{2/3}\,.
\label{divergence}
\ee
Thus, from (\ref{pdTP}), \textit{any physical quantity given by the temperature derivative along the critical isobar  diverges  with the  universal  exponent $2/3$}. This includes the critical exponent of the specific heat per particle under constant pressure:
\be
\ba{ll}
C_{P}~\sim~{\left|{T/T_{c}-1}\right|}^{-\alpha}\,,~~~&~~~\alpha=2/3\,.
\ea
\label{alphaP}
\ee
On the other hand,  since  the specific heat per particle at constant volume is finite for any finite system,  the corresponding critical exponent is trivial:
\be
\ba{lll}
C_{V}\,=\,\mbox{finite}~~&~~\mbox{\textit{i.e.}}~~&~~\alpha=0\,.
\ea
\label{alphaV}
\ee
Note that our conclusion does not exclude the possibility for $C_{V}$ to develop a   peak. We merely point out that  the peak will be smoothed out if it is sufficiently zoomed in.  
Further from (\ref{divergence}), the inverse of $\Phi$ gives the critical exponent of the isothermal compressibility on the critical isobar:
\be
\ba{ll}
\kappa_{{\scriptscriptstyle{T}}}:=-V^{-1}\partial_{P}V(P_{c},T)~\sim~{\left|{T/T_{c}-1}\right|}^{-\gamma}\,,~~&~~\gamma=2/3\,.
\ea
\label{gamma}
\ee
Finally, from (\ref{criticalprop1}), we also obtain an \textit{isothermal} critical exponent at the critical temperature, ${T=T_{c}}\,$:
\be
\ba{ll}
P/P_{c}-1~\sim~{\left|{V/V_{c}-1}\right|}^{\delta}\,,~~~&~~~\delta=3\,.
\ea
\label{delta}
\ee
~\\

\begin{figure}[H]
\normalsize
\includegraphics[width=81mm]{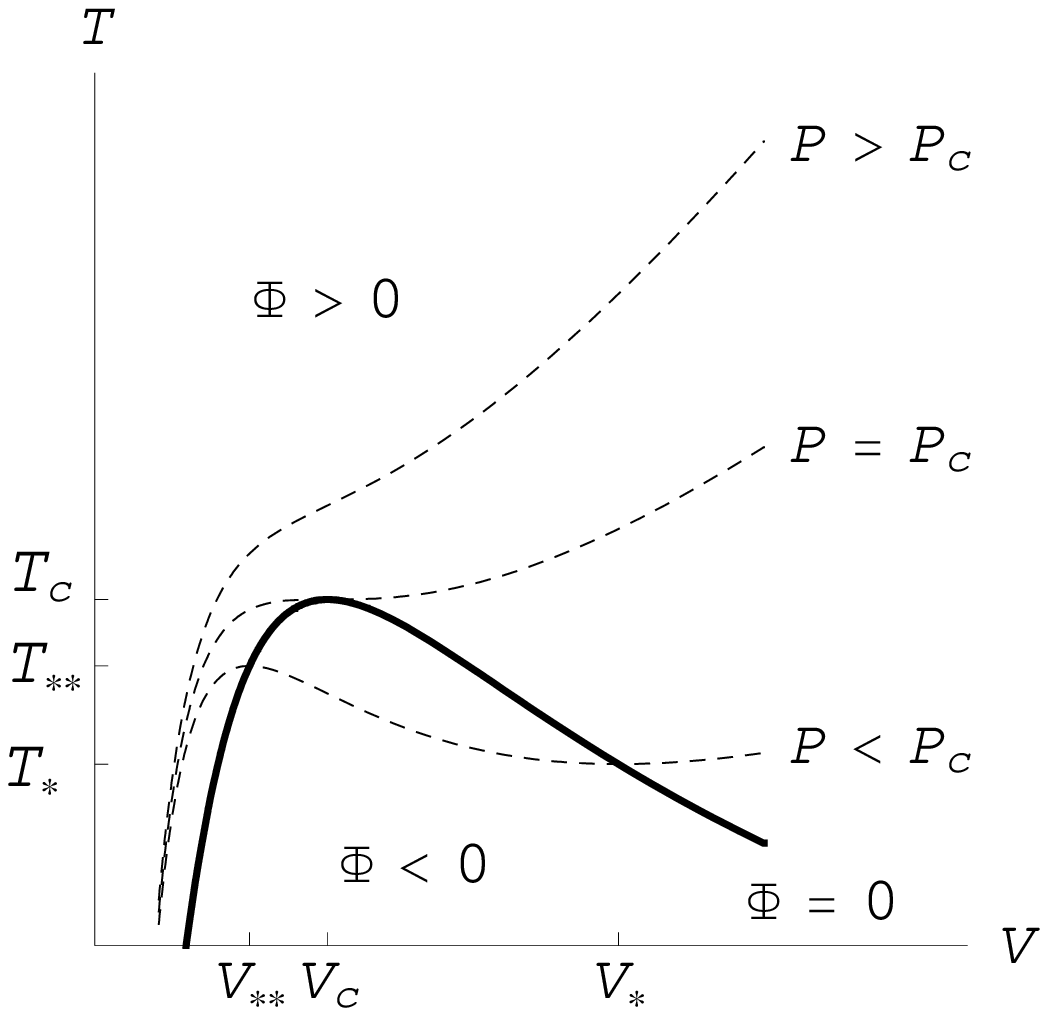}
\caption[]{{\textbf{Van der Waals equation of state}} \\
{\sf The  dashed lines are constant pressure lines of three different values, and  the thick  solid line is the spinodal curve   satisfying ${\Phi=-\partial_{V}P(T,V)=0}$.  When ${P<P_{c}}$
  the constant pressure line crosses  the spinodal curve  twice, at the supercooling point ${(V_{\ast},T_{\ast})}$ and at the  superheating point ${(V_{\ast\ast},T_{\ast\ast})}$. Between the two temperatures the volume is triple valued.  If ${P=P_{c}}$ the constant pressure line comes in contact with the spinodal  curve only once at the critical point ${(V_{c},T_{c})}$. Otherwise, \textit{i.e.~}${P>P_{c}}$, the constant pressure line does not undergo thermodynamic instability and hence no first-order phase transition arises.  On the spinodal curve the critical point has the highest temperature. }}
\label{relativisticFIGVanderWaals}
\end{figure}

 On a generic  isobar below the critical pressure, ${P<P_{c}}$,  the thermodynamic instability given by   ${\Phi(P,V)<0}$ appears  naturally  over an interval in volume and hence in  temperature as well.  The  two ends then  correspond to  a supercooling point and a superheating point.  At these points  $\Phi$ vanishes and between them $\Phi$ is negative, as depicted   in  FIG.~\ref{relativisticFIGVanderWaals} from the Van der Waals equation of state as an example:\footnote{For a modern exposition of Van der Waaals forces, see \textit{e.g.}~\cite{vdWBOOK}.}
\be
\textstyle{\left[(P/P_{c})+3(V_{c}/V)^{2}\right]}\textstyle{[(V/V_{c})-\textstyle{\frac{1}{3}}]} =\textstyle{\frac{8}{3} (T/T_{c})}\,.
\ee
It is crucial to note that  the volume   assumes   triple  values between the supercooling and the superheating temperatures at  constant  pressure less than $P_{c}\,$.\\ 

\noindent

In the derivation of the critical exponents above we tacitly  assumed  $\partial_{V}^{2}\Phi(P_{c},V_{c})\neq 0$.
In fact, the equivalence among (\ref{criticaldef}), (\ref{criticalprop1}) and (\ref{criticalprop2}) generalizes  up to an arbitrary order, say $n\,$: For all $k=1,2,\cdots,n$, the vanishing of
 $\partial^{k}_{V}P(T,V)$ is equivalent to that of  $\partial_{V}^{k}T(P,V)$
  as well as that of  $\partial^{k-1}_{V}\Phi(P,V)$.  Hence, if the first non-vanishing multiple volume derivative of $\Phi$ occurs at the order ${n}$ as  $\partial^{n}_{V}\Phi(P_{c},V_{c})\neq 0$,  the exponents in (\ref{beta}), (\ref{alphaP}), (\ref{gamma}), (\ref{delta}),  assume the general values:
\be
\ba{ll}
\alpha=\gamma=\textstyle{\frac{n}{n+1}}\,,~~~&~~~~\beta=\delta^{-1}=\textstyle{\frac{1}{n+1}}\,.
\ea
\label{generalexponent}
\ee
Clearly these satisfy the following ``scaling laws"~\cite{Huang,Stanley}:
\be
\ba{rl}
\mbox{Rushbrooke}:~~&~~\alpha+2\beta+\gamma=2\,,\\
\mbox{Widom}:~~&~~\gamma=\beta(\delta-1)\,.
\ea
\ee
Especially, the case of  ${n=1}$ corresponds to the first-order phase transition at noncritical, supercooling and superheating points. Even at these points our analysis shows the existence of the universal exponent:
\be
\alpha=\beta=\gamma=\delta^{-1}=\half\,,
\label{exponentnequalone}
\ee
which  agrees with  the  mean field theory result~\cite{MEANFIELD} (see also \cite{Speedy}). 
Around the supercooling or  superheating point  in FIG.~\ref{relativisticFIGVanderWaals}, the constant pressure lines can be approximated by a parabola straight up or down respectively. The possibilities of  $n\geq 3$ shall correspond to exceptional critical points.  \\

In the next section we show that ideal relativistic quantum gas of neutral bosons  indeed features  a spinodal curve similar to   FIG.~\ref{relativisticFIGVanderWaals}. \newpage

\section{Ideal relativistic neutral Bose gas\label{secBose}}
\subsection{Algebraic analysis}
Hereafter,   exclusively  for  ideal relativistic neutral Bose gas,  we focus on $N$  particles with mass $m$, confined  in a  box  of dimension $d$ and length $L\equiv V^{1/d}$.   Hard, impenetrable walls impose Dirichlet boundary condition.\footnote{We recall that nevertheless enforcing  periodic or Neumann boundary condition leads to  thermodynamic instability at low temperature near to absolute zero for arbitrary $N$~\cite{7616Park}.}

With  \textit{positive} integer valued good quantum numbers:
\be
\vec{n}=(n_{1},n_{2},\cdots,n_{d\,})\,,
\ee
the spatial momentum is quantized as $\vec{p}=\pi\hbar\vn/L$, such that  the single particle Boltzmann factor in (\ref{lambda})  assumes the form:
\be
\textstyle{e^{-\beta E_{\vec{n}}}=\exp\!\left({-\,\cT^{-1}\sqrt{1+\vn{\cdot\vn}\,\cV^{-2/d}\,}}\,\right).}
\label{Boltzmann}
\ee
Here  we have introduced, for  simplicity of notation and the forthcoming analysis,   \textit{dimensionless temperature}:
\be
\dis{\cT:=\kB T/(mc^{2})\,,}
\label{dtemperature}
\ee
as well as  \textit{dimensionless volume}:
\be
\dis{\cV:=[mc/(\pi\hbar)]^{d\,}V\,.}
\label{dvolume}
\ee
The canonical partition function is then two-variable  function of them, $Z_{N}(\cT,\cV)$. 
We further  define \textit{dimensionless pressure} by:
\be
\cP:=[(\pi\hbar)^{d}/(m^{d+1}c^{d+2})]\,P=\cT\partial_{\cV}\ln Z_{N}(\cT,\cV)\,,
\label{dpressure}
\ee
and  a dimensionless indicator of the thermodynamic instability, as for $\Phi$ in (\ref{Phi}):
\be
\,\phi\,\,:=-(1/N)\cV^{2}\partial_{\cV}^{2}\ln Z_{N}(\cT,\cV)\,.
\label{phi}
\ee
Other physical quantities we are particularly   interested in the present paper are 
specific heats per particle  at constant volume or under constant pressure:
\be
\ba{l}
{C_{V}=(\kB/N)\left[(\cT\partial_{\cT})^{2}+\cT\partial_{\cT}\right]\ln Z_{N}(\cT,\cV)\,,}\\
{C_{P}=C_{V}-\kB\cT[\partial_{\cT}\cP(\cT,\cV)]^{2}/[N\partial_{\cV}\cP(\cT,\cV)]\,.}
\ea
\ee
~\\

\noindent $\bullet$ \textit{Non-relativistic   limit.} The  non-relativistic limit corresponds to the limit of  low temperature and large volume with $\cT\cV^{2/d}$ held fixed.  For large  volume,  the Boltzmann factor (\ref{Boltzmann}) reduces to that of the non-relativistic  particle, up to the  shift of the  energy by $mc^{2}$:
\be
\textstyle{e^{-\beta E_{\vec{n}}}\rightarrow e^{-\beta[mc^{2}+\vec{p}{\cdot\vec{p}}/(2m)]}=
e^{-{1/\cT}-\vn{\cdot\vn}/(2\cT\cV^{2/d})}\,.}
\ee
Physical quantities like  $\phi$, $C_{V}$, $C_{P}$   become one-variable  functions depending  on the single variable $\cT\cV^{2/d}$,  or alternatively $\cT\cP^{-2/(d+2)}$ \cite{7616Park}. The constant energy shift is irrelevant to them. Furthermore, at extremely  low temperature only the ground state energy matters and   the canonical partition function reduces to
\be
\textstyle{Z_{N}\rightarrow e^{-\beta NE_{0}}=\exp\!\left({-\,N\cT^{-1}\sqrt{1+d\,\cV^{-2/d}\,}}\,\right).}
\label{Zatzero}
\ee
Hence near to absolute zero temperature we obtain 
\be
N/\cP=\cV^{1+1/d}\sqrt{d+\cV^{2/d}\,}\,,
\label{lowtempP}
\ee
and also
\be
\ba{lll}
{\phi=\infty}\,,~~~&~~~~{C_{V}=0\,,}~~~&~~~~{C_{P}=0\,.}
\ea
\label{zerotemp}
\ee
For large volume,  Eq.(\ref{lowtempP}) further leads to $\cV=(N/\cP)^{d/(d+2)}$ and this agrees with Ref.\cite{7616Park}.  Apparently the volume assumes a finite value at absolute zero temperature, essentially    due to the Heisenberg uncertainty principle. It is worthwhile to recall
\be
\ba{llll}
{PV=N\kB T\,,\!}&{\phi=1\,,\!}&{C_{V}{/\kB}={d/2}\,,\!}&{C_{P}{/\kB}=1+{d/2}\,,\!}
\ea
\label{nonrellimit}
\ee
in the   large  $\cT\cV^{2/d}$  limit  of  the non-relativistic case~\cite{7616Park}. Note that the first relation is equivalent to $\cP\cV=N\cT$. \newline

\noindent $\bullet$ \textit{Ultra-relativistic   limit.} The ultra-relativistic limit corresponds to the limit of  high temperature and small volume with $\cT\cV^{1/d}$ held fixed.   At sufficiently  high temperature,   highly excited  states dominate and from (\ref{conapp}) the canonical partition function reduces to $(Z_{1})^{N}/N!$, where  the Boltzmann factor  also reduces as
\be
\textstyle{e^{-\beta E_{\vec{n}}}\longrightarrow e^{-\beta c|\vec{p}|}=
\exp\!\left[-|\vec{n}|/(\cT\cV^{1/d})\right].}
\label{ultraBoltzmann}
\ee
Consequently,  the canonical partition function and the physical quantities like  $\phi$, $C_{V}$, $C_{P}$ become  one-variable  functions  depending on  the single variable $\cT\cV^{1/d}$, or alternatively from (\ref{dpressure}) and (\ref{ultraBoltzmann})  on $\cT\cP^{-1/(d+1)}$.

Moreover, for large $\cT\cV^{1/d}$  we may safely assume the continuous approximation to obtain~\cite{Blundell},
\be
Z_{N}(\cT,\cV)\propto (\cV\cT^{d})^{N}\,,
\ee 
which  implies in the  large $\cT\cV^{1/d}$ limit, \textit{cf.} (\ref{nonrellimit}):
\be
\ba{llll}
{PV=N\kB T\,,}~&{\phi=1\,,}~&{C_{V}/\kB=d\,,}~&{C_{P}/\kB=1+d\,.}
\ea
\label{ultralimit}
\ee
~\newline
\subsection{Numerical results\label{sectionNUMERICAL}}
Here we present our numerical results on the ideal relativistic neutral Bose gas confined in a cubic box,  \textit{i.e.~}${d= 3}$,  at  generic temperature and volume, performed by   a supercomputer  ({\small SUN B6048}).  Our  analysis  is based on the    recurrence relation   (\ref{recZ}) and  the  previous work~\cite{7616Park}\footnote{The numerical computation inevitably requires a  momentum  cutoff as for  $\vn{\cdot\vn}$, which generically deforms the result at   high temperature.  We choose mutually different,  sufficiently large  values of the cutoff, such that we maintain  at least for  some interval of temperature  the high temperature behaviour  (\ref{ultralimit}),   and report   only the  cutoff independent  results (\textit{c.f.~}Refs.\cite{Glaum,Kleinert}).  }. \\

\begin{figure}[H]
\normalsize
\includegraphics[width=104mm]{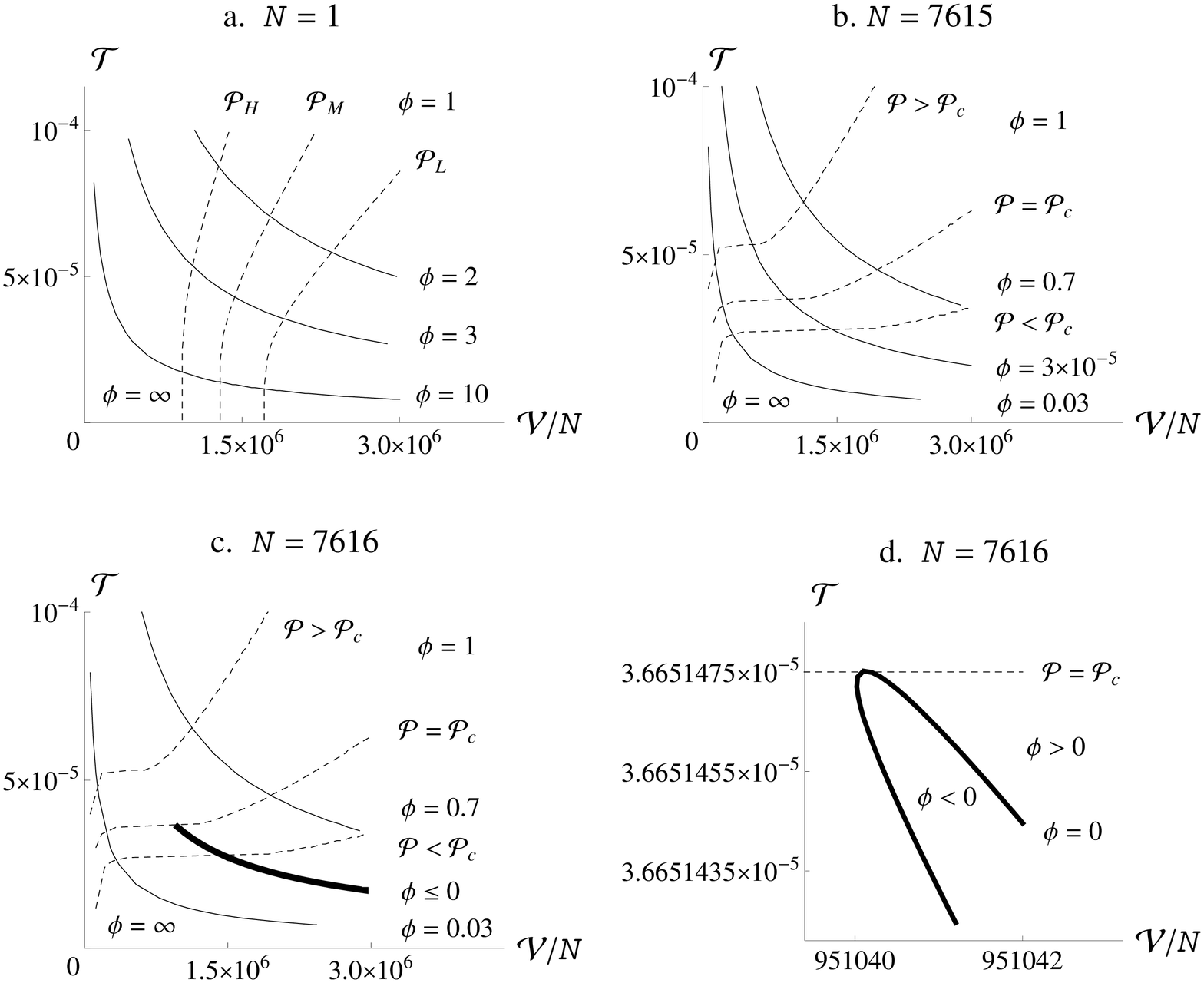}
\caption[]{{\textbf{Constant pressure lines and spinodal curve}} \\
{\sf Dashed, thin solid or thick solid lines denote respectively  the constant pressure, constant $\phi$ or spinodal curve. Near to the origin of the $(\cT,{\cV/N})$ plane $\phi$ diverges and in the opposite  infinite limit $\phi$ converges to unity. When ${N=1}$ (FIG.\,\ref{relativisticFIGphi}\textbf{a} with $\cP_{L}<\cP_{M}<\cP_{H}$), $\phi$ is monotonically decreasing from $\infty$ to $1$ on arbitrary isobars. As $N$ increases $\phi$ develops a valley whose height  is  less than unity. Moreover if $N\geq 7616$  the valley assumes negative values and a spinodal curve emerges. FIG.\,\ref{relativisticFIGphi}\textbf{d} magnifies the tip of the spinodal curve for ${N=7616}$ to manifest a critical point,  $(\cT_{c},{\cV_{c}/N})=({3.665\,1475\times 10^{-5}},\,{9.510\,401\times 10^{5}})$.   In FIG.\,\ref{relativisticFIGphi}\textbf{b},  $\cP_{c}$ denotes  merely    the numerical value,  $\cP_{c}=2.015\,1967\times 10^{-11}$, which amounts to the critical pressure in the system with one more particle,  $N=7616$.}}
\label{relativisticFIGphi}
\end{figure}

\noindent $\bullet$ \textit{Emergence of a spinodal curve}  (FIG.\,\ref{relativisticFIGphi}). As $N$ grows, $\phi$ develops a valley of local minima, which eventually assumes   negative values if $N\geq 7616$.\\

\noindent

\newpage

\begin{figure}[H]
\normalsize
\includegraphics[width=123mm]{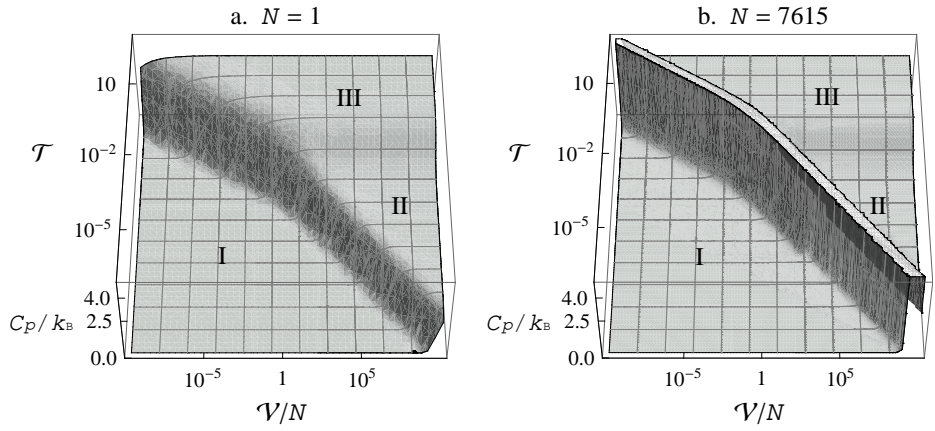}
\caption[]{{\textbf{$3D$ image of $C_{P}$ on $(\cT,{\cV/N})$ plane for ${N=1}$ and ${N=7615}$ }} \\
{\sf The plane decomposes into three parts: phase $\mathbf{I}$ with ${C_{P}\simeq 0}$, phase $\mathbf{II}$ with ${C_{P}\simeq 2.5\kB}$ and phase $\mathbf{III}$ with ${C_{P}\simeq 4\kB}$. 
When ${N=1}$ the transitions are monotonic and smooth. As $N$ grows, at  the borders  between $\mathbf{I}$  and  $\mathbf{II}$ as well as  between  $\mathbf{I}$ and  $\mathbf{III}$,  a range of peaks emerges which will eventually diverge for ${N\geq 7616}$. FIG.\,\ref{relativisticFIG3D}$\textbf{b}$ has  been cut at the height of $C_{P}=5\kB$, and  the actual peak rises up to $C_{P}\simeq 5.336\,84\times10^{6}
\kB$.  }}
\label{relativisticFIG3D}
\end{figure}

\begin{figure}[H]
\normalsize
\includegraphics[width=128mm]{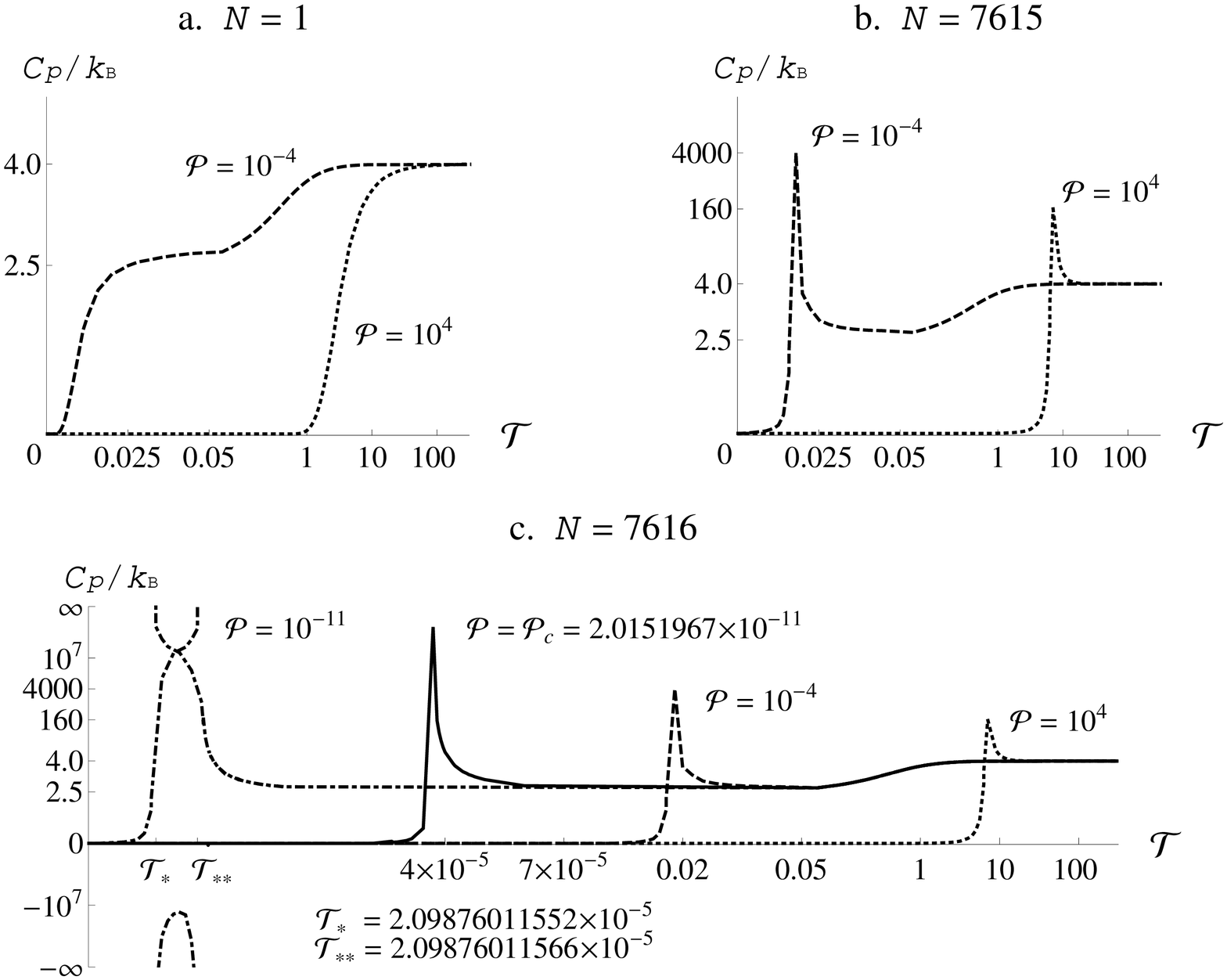}
\caption[]{{\textbf{$C_{P}$ for various $N$ and $\cP$}} \\
{\sf
For ${N=1}$, $C_{P}$ is monotonically  increasing from zero to $4\kB$ on any isobar. Under sufficiently low pressure, it assumes the intermediate value of $2.5\kB$ as in Eq.(\ref{nonrellimit}). As $N$ increases, $C_{P}$ develops a peak on each isobar. Especially when ${N\geq 7616}$ and ${\cP<\cP_{c}}$, it diverges both to the plus and minus infinities at the supercooling point ${\cT=\cT_{\ast}}$ 
 as well as at the superheating point ${\cT=\cT_{\ast\ast}}$. At the critical point  ${\cT=\cT_{c}}={3.665\,1475\times 10^{-5}}$, it diverges  only positively. For ${\cP>\cP_{c}}$,  the specific heat  features a single finite peak which can be identified as  the Widom line.}}
\label{relativisticFIGcp}
\end{figure}

\begin{figure}[H]
\normalsize
\includegraphics[height=139mm]{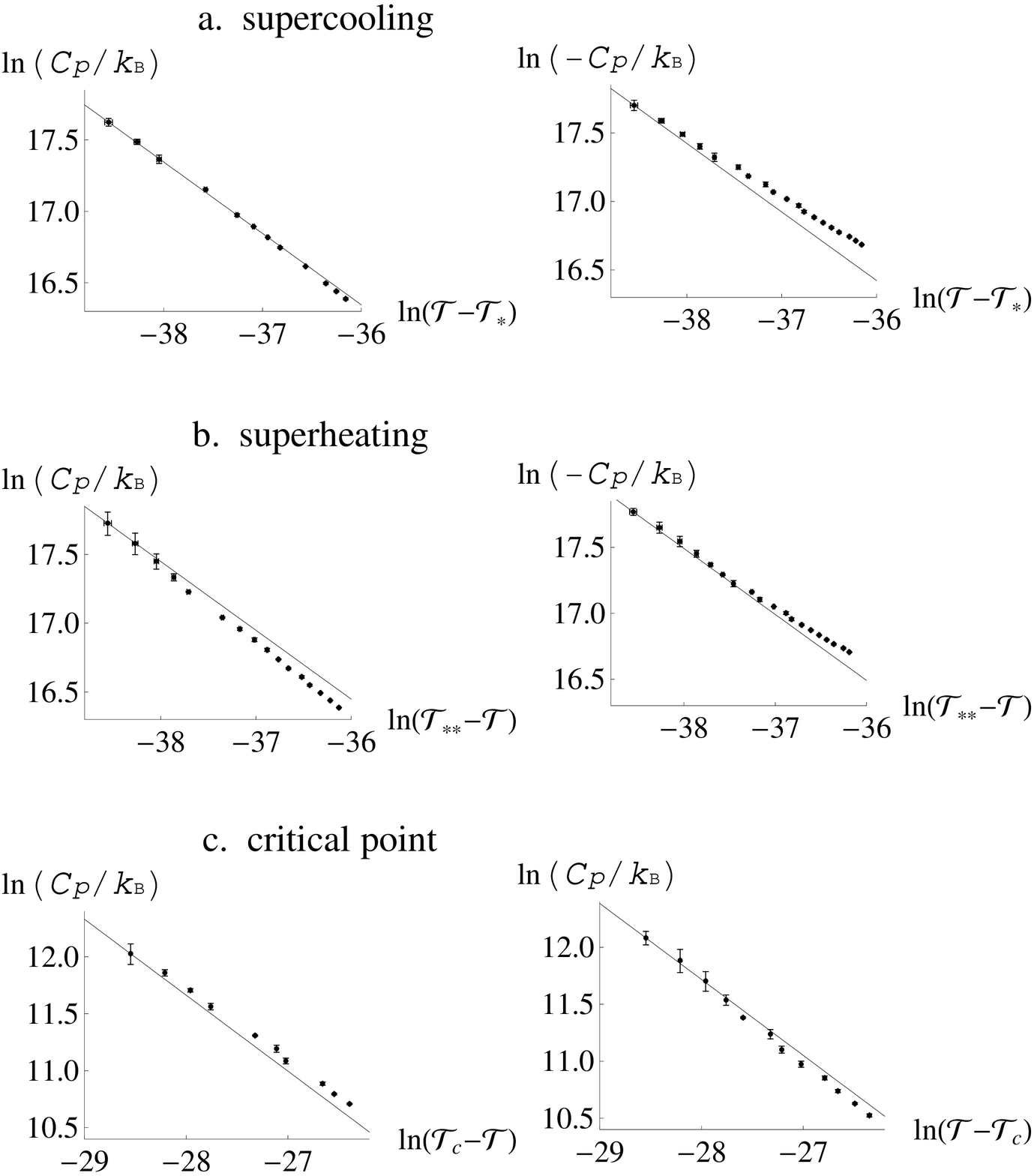}
\caption[]{{\textbf{Critical and noncritical exponents}} \\
{\sf  Our numerical data confirm the singular behavior of the specific heat $C_{P}$ anticipated in Sec.\,\ref{Secexponent}:  above  the supercooling point,  
$C_{P}\sim\pm (\cT-\cT_{\ast})^{-1/2}$ (FIG.\,\ref{relativisticFIGexponents}{\bf a});  below the superheating point,   
$~C_{P}\sim\pm (\cT_{\ast\ast}-\cT)^{-1/2}$  (FIG.\,\ref{relativisticFIGexponents}{\bf b}); and     around the  critical point,   $~C_{P}\sim |\cT-\cT_{c}|^{-2/3}$ (FIG.\,\ref{relativisticFIGexponents}{\bf c}). The numerical data are for ${N=7616}$, $\cP=5.000\,170\,056\,40\times 10^{-12}$ (FIG.\,\ref{relativisticFIGexponents}{\bf a,b}) or $\cP_{c}=2.015\,1967\times 10^{-11}$ (FIG.\,\ref{relativisticFIGexponents}{\bf c}).  The error bars originate from   the numerical uncertainty  beyond  these digits.    The straight lines correspond to the theoretical slopes, $-{1/2}$ (FIG.\,\ref{relativisticFIGexponents}{\bf a,b}) or $-{2/3}$ (FIG.\,\ref{relativisticFIGexponents}{\bf c}).  }}
\label{relativisticFIGexponents}
\end{figure}

\noindent $\bullet$ \textit{Specific heat per particle under constant pressure} (FIGs. \ref{relativisticFIG3D}, \ref{relativisticFIGcp}, \ref{relativisticFIGexponents}). During the first-order phase transition under constant pressure which is less than the critical value, the volume and hence every physical quantity are  triple valued between the supercooling and the superheating temperatures.  We focus here on $C_{P}$, the specific heat per particle under constant pressure. For the explicit behavior  of  other quantities  we refer to our earlier work~\cite{7616Park}.\\

\noindent


\newpage

\begin{figure}[H]
\normalsize
\includegraphics[width=120mm]{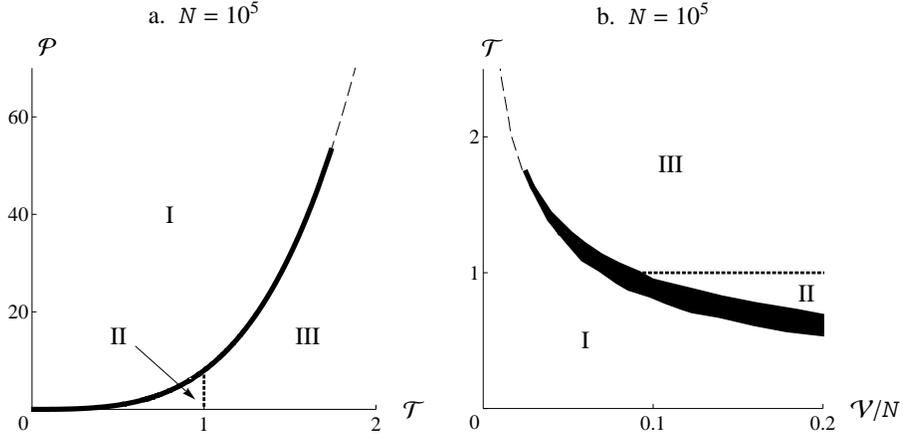}
\caption[]{{\textbf{Phase diagram for $N=10^{5}$ on $(\cP,\cT)$ and $(\cT,{\cV/N})$ planes}} \\
{\sf Thick solid line,  dashed line and dotted line correspond to the spinodal curve (including its inside), the Widom line (a range of finite peaks in $C_{P}$  which also coincides with the valley of $\phi$)  and the $\cT\simeq 1$ line respectively. The lines divide the  phase diagram  into  three parts: phase $\mathbf{I}$ of ${C_{P}\simeq 0}$, phase $\mathbf{II}$ of  ${C_{P}\simeq 2.5\kB}$ and phase $\mathbf{III}$ of ${C_{P}\simeq 4\kB}$.  On the top of the solid line, there exists a critical point. As $N$ increases, the critical point moves  toward more ultra-relativistic,  higher temperature  region along the Widom line  keeping $\cT_{c}(\cV_{c}/N)^{1/3}$ and $\cT_{c}\cP_{c}^{-1/4}$ constant.  }}
\label{relativisticFIGphasediagram}
\end{figure}

\noindent $\bullet$ \textit{Phase diagram of the ideal relativistic neutral Bose gas   for  $N=10^{5}$.} (FIG.\,\ref{relativisticFIGphasediagram}).

\newpage

Our numerical results  of  the critical point  are listed below for  some selected  $N$:
{\small{\begin{center}
\begin{tabular}{l|l|l|l}
{$N$}&$\cT_{c}$&$\cV_{c}/N$&$\cP_{c}$  \\
\hline
$7616$&$3.665\,1475\!\times\! 10^{-5}$&$9.510\,401\!\times\! 10^{5}$&$~\,2.015\,1967\!\times\! 10^{-11}$\\
$10^{4}$&$0.149\,54$&$ 3.133$&$~\,0.027\,02$\\
$10^{5}$&$1.738$&$0.025\,2$&$53.1$\\
\end{tabular}
\end{center}}}
~\newline
which gives
{\small{\begin{center}
\begin{tabular}{l|l|l|l|l}
{$N$}&$\cT_{c}(\cV_{c}/N)^{2/3}$&$\cT_{c}\cP_{c}^{-2/5}$&$\cT_{c}(\cV_{c}/N)^{1/3}$&$\cT_{c}\cP_{c}^{-1/4}$~~  \\
\hline
$7616$&$0.354\,4519$&$0.695\,6079$&$3.604\,329\times 10^{-3}$&${1.729\,865}\times 10^{-2}$\\
$10^{4}$&$0.320\,2$&$0.634\,0$&$0.218\,8$&$0.368\,8$\\
$10^{5}$&$0.149$&$0.355$&    $0.510$& $0.644$\\
\end{tabular}
\end{center}}}
~\newline
Comparison  with the non-relativistic results  in Ref.\cite{7616Park} indicates that  the critical point of ${N=7616}$ is in the non-relativistic region while those of $N=10^{4}$ and $N=10^{5}$ are not.\newpage

\section{Summary and comments \label{secDISCUSSION}}
In summary, as seen from FIG.~\ref{relativisticFIG3D} and  FIG.~\ref{relativisticFIGphasediagram}, the ideal relativistic neutral Bose gas divides  the $(\cT,{\cV/N})$ phase diagram  into three parts.
\begin{itemize}
\item  Phase $\mathbf{I}$: \textit{condensate} with  ${C_{p}\simeq 0}$ (\ref{zerotemp}),
\[\left\{~(\cT,{\cV/N})~~|~~ \cT\lesssim ({N/\cV})^{1/3}~\mbox{\&}~\cT\lesssim ({N/\cV})^{2/3}\right\}.\]
\item   Phase $\mathbf{II}$: \textit{non-relativistic gas} with   ${C_{p}\simeq  {2.5\kB}}$ (\ref{nonrellimit}),
\[\left\{~(\cT,{\cV/N})~~|~~ \cT\lesssim 1~\mbox{\&}~\cT\gtrsim ({N/\cV})^{2/3}\right\}.\]
\item    Phase $\mathbf{III}$: \textit{ultra-relativistic gas}  with  $C_{p}\simeq  {4\kB}$  (\ref{ultralimit}),
\[\left\{~(\cT,{\cV/N})~~|~~ \cT\gtrsim ({N/\cV})^{1/3}~\mbox{\&}~\cT\gtrsim 1~\right\}.\]
\end{itemize}
Equivalently the $(\cP,\cT)$ phase diagram splits  into three parts.
\begin{itemize}
\item Phase $\mathbf{I}$: $\mbox{~~}\textstyle{ \left\{~(\cP,\cT)~|~ \cT\lesssim \cP^{2/5}~\mbox{\&}~\cT\lesssim\cP^{1/4}\right\}}$.
\item Phase $\mathbf{II}$:  $\mbox{~}\textstyle{\left\{~(\cP,\cT)~| ~\cT\gtrsim \cP^{2/5}~\mbox{\&}~\cT\lesssim 1\right\}}$. 
\item Phase $\mathbf{III}$:  $\textstyle{\left\{~(\cP,\cT)~|~\cT\gtrsim \cP^{1/4}~\mbox{\&}~\cT\gtrsim 1\right\}}$.
\end{itemize}
Here $\lesssim$ and $\gtrsim$ mean rough inequalities up to  constants of order unity.\\

When $N=1$, the transitions are all smooth and monotonic along any isobar.  As $N$ increases,  a valley  in $\phi$ and hence a range of peaks in $C_{P}$ develops  along the boundary of the  condensate phase $\mathbf{I}$.  
Especially, when ${N=7616}$, the valley of $\phi$ assumes  negative values and a spinodal curve with a critical point emerges along the  boundary of the phase  $\mathbf{I}$ starting from the absolute  zero temperature.  Beyond the critical point it is the Widom line that divides the phase $\mathbf{I}$ and the phase $\mathbf{III}$.  As $N$ further increases, the critical point moves   along the Widom line toward  the ultra-relativistic higher temperature region.   For the transition  between the non-relativistic and ultra-relativistic gas phases  $\mathbf{II}$ and  $\mathbf{III}$ there is  no latent heat involved. \\

The spinodal curve sharply defines the phase diagram. The consequent  phase transition  is  first-order below the critical pressure  or second-order at the    pressure. The   exponents  corresponding to the singularities are   $1/2$  and $2/3$  respectively. Presence of both the supercooling and the superheating characterizes the first-order phase transition.\\

The resulting equation of state from  the spinodal curve  of the relativistic ideal gas resembles the Van der Waals equation of state which is derived by assuming both  repulsive potential at short distance and attractive potential at long distance.  It is well known that the effective  statistical interaction of ideal Bose gas is attractive~\cite{Huang}.  
Our result seems to suggest a more intriguing structure. The fact that the volume is finite  at absolute zero temperature seems  also to indicate a repulsive interaction at short distance related to the Heisenberg uncertainty principle. \\

An interesting open question is whether   the critical point converges or not in the phase diagram as $N\rightarrow\infty$.  If not, one may also wonder whether there exists another critical number in $N$ for the emergence of a spinodal curve from the  ultra-relativistic canonical partition function (\ref{ultraBoltzmann}).  We have verified  for up to $N=10^{6}$ that the ultra-relativistic canonical partition function does not feature any spinodal curve, though $\phi$ develops a local minimum.  \\

Although in this work we have focused on the relativistic generalization of the non-relativistic ideal Bose gas   and have obtained a nontrivial  phase diagram with a critical point and Widom line,  it is natural to expect that other generalizations which involve extra dimensionful  parameters, \textit{e.g.}  trapping potentials, may also lead to qualitatively similar or even more intriguing   phase diagrams.\\  \newpage

\noindent\textbf{Acknowledgements}  We  thank  Imtak Jeon,   Konstantin Glaum,  Hagen Kleinert  and Giovanna Simeoni for useful comments.  This work was supported by the National Research Foundation of Korea(NRF) grant funded by the Korea government(MEST) with the grant numbers 2005-0049409 (CQUeST),  2009-352-C00015 (I00216)  and 2010-0002980.\\


\newpage





\end{document}